\documentclass{A-iopjournal}

\usepackage[T1]{fontenc}
\usepackage{lmodern}
\usepackage[numbers,sort&compress]{natbib}%
\usepackage{graphicx}%
\usepackage{multirow}%
\usepackage{amsmath,amssymb,amsfonts}%
\usepackage{amsthm}%
\usepackage{mathrsfs}%
\usepackage[title]{appendix}%
\usepackage{xcolor}%
\usepackage{textcomp}%
\usepackage{manyfoot}%
\usepackage{booktabs}%
\usepackage{algorithm}%
\usepackage{algorithmicx}%
\usepackage{algpseudocode}%
\usepackage{listings}%
\usepackage{enumitem}%
\usepackage{xcolor}

\newtheorem{theorem}{Theorem}%
\newtheorem{criteria}{Criterion}
\newtheorem{definition}{Definition}%

\begin{document}

\articletype{Paper} 

\title{The Twin Paradox in the Vicinity of Rotating Black Holes}

\author{Shuiquan Bai$^*$\orcid{0009-0009-2644-4252} and Geraint F. Lewis\orcid{0000-0003-3081-9319}}

\affil{Sydney Institute for Astronomy, School of Physics, A28, The University of Sydney, NSW 2006, Australia}

\affil{$^*$Author to whom any correspondence should be addressed.}

\email{sbai0925@uni.sydney.edu.au}

\keywords{relativity, twin paradox, black holes, Jacobi field}

\begin{abstract}
The twin paradox is a foundational thought experiment in the special theory of relativity where a returning twin ages less than the one who remains stationary. However, the intricacies of the twin paradox remain relatively underexplored in the curved spacetimes of general relativity. Here we explore the twin paradox in the vicinity of a rotating black hole, where the existence of multiple paths between two events creates significant complexity. We develop a numerical framework based on residual maps and optimisation to identify possible trajectories. We find a strong negative correlation between a traveller's experienced proper time and both the azimuthal distance travelled and the magnitude of acceleration. We apply numerical Jacobi field analysis to examine conjugate points along geodesics within the Kerr geometry, finding that only the geodesic with the minimal azimuthal distance contains no conjugate points. This provides beginner students of general relativity with a visual tool to understand general relativistic concepts, helping to correct flat-spacetime intuitions.
\end{abstract}

\section{Introduction}\label{chapter:intro}
For over a century, Einstein’s theory of relativity has been a cornerstone of modern physics \cite{einstein1905elektrodynamik,einstein1916}. One of the most famous thought experiments is the twin paradox, which has long been a valuable tool for teaching and understanding relativistic concepts. In flat spacetime, a twin travelling on an interstellar journey returns, due to time dilation, younger than the twin remaining on Earth. While the principle of relativity suggests the traveller could view the Earth twin as moving, this poses no logical contradiction. The standard answer is that the accelerated twin is always younger \cite{langevin1911evolution, von1913relativitatsprinzip, minkowski1909raum, Builder_1959}. Among these, Schild \cite{schild1959clock} reviews the history of the paradox and its common misconceptions, and provides a very clear geometric account in special relativity, showing through Minkowski geometry why the inertial worldline corresponds to the maximum proper time.

Extending the twin paradox into curved spacetime increases complexity, as multiple geodesics between two events can exist. In such cases, the geodesic condition alone does not guarantee maximal proper time. Tangherlini \cite{tangherlini1962postulational} was among the first to demonstrate this explicitly by showing in a uniform-density interior Schwarzschild solution that the standard conclusion from special relativity may be reversed. One particularly interesting situation is with black holes. Black hole metrics often serve as the core spacetimes studied in general relativity teaching. However, in these environments, the simple flat-spacetime conclusion, that the accelerated twin is always younger, no longer holds. Although the twin paradox has been investigated across various curved spacetimes \cite{holstein1972relativity,markley1973relativity,abramowicz2007twin,lichtenegger2011twin,boblest2011twin,gron2011twin,gron2013twin}, research involving black holes has primarily focused on the Schwarzschild geometry. For example, Abramowicz and Bajtlik \cite{abramowicz2009adding} demonstrated that an accelerated twin can actually be older than a twin in free fall orbiting the Schwarzschild spacetime, a result further explored through numerical methods by Fung et al. \cite{fung2016computational}. Yet, these studies have not directly provided a clear induction of what the flat-spacetime intuition becomes in the vicinity of rotating black holes. This leads to a conceptual gap for beginners in general relativity, hindering their intuitive understanding of the age differences in a curved spacetime.

To bridge this gap, we establish a numerical framework to resolve the multiple timelike paths connecting two events within the Kerr geometry. Furthermore, we introduce numerical Jacobi field analysis to examine conjugate points along geodesics, providing a criterion for evaluating proper-time maximisation.

The paper is structured as follows: In Section \ref{sec:eduframe}, we outline the educational framework, detailing the target levels and the intended learning objectives. In Section \ref{sec:methods}, we present the equations of motion and detail our numerical framework, covering both the residual analysis procedure and the foundation of Jacobi fields. In Section \ref{sec:results}, we then report the numerical results and highlight the dependence of proper time on acceleration and azimuthal distance. In Section \ref{sec:implications}, we discuss the instructional implementation across different stages of a general relativity curriculum, before summarising key findings in Section \ref{sec:conclusion}.


\section{Educational Framework}\label{sec:eduframe}
The primary goal of this work is to build up general relativity intuition through numerical exploration. Rather than deriving analytical solutions, which are often intractable in complex scenarios, the approach focuses on constructing the equations of motion from spacetime metrics for numerical integration. The auxiliary quantities, such as Christoffel symbols, are obtained from the Catalogue of Spacetimes \cite{mueller2010cataloguespacetimes}, allowing the focus to remain on the exploration of the underlying physics.

To structure the instructional applications, the target students are divided into two categories. For beginner students of general relativity starting to explore black hole metrics, the numerical framework combines optimisation with residual maps to offer a direct visual intuition. Inspired by the concept of winding number in topology \citep[see e.g.][]{uzan2002twin, bansal2005twin} and gravitational lensing image classification \cite{PhysRevD.62.084003}, azimuthal distance is used as a second indicator of proper time to classify different trajectories and provide straightforward images for beginners. This classification serves as a conceptual bridge to show what the flat-spacetime intuition becomes in the vicinity of rotating black holes. Students with Python experience can reproduce these numerical simulations through computational assignments to verify the physical behaviours for a deeper understanding.

For advanced or graduate students who study the theory of Jacobi fields, the existence of conjugate points provides a criterion to determine whether an extended timelike path connecting two distinct events constitutes a local maximum of proper time \cite{Hawking_Ellis_1973, Wald:1984rg, do2016differential, do1992riemannian, milnor1963morse}. While the Jacobi field has been applied to simple cases \cite{sokolowski2012twin,sokolowski2014twin,sokolowski2015jacobi,sokolowski2017geometric}, the complexity of the Jacobi equation limits the scope of analytical solutions. The numerical Jacobi field analysis presented here translates these abstract concepts into computational results, providing a practical framework to identify geodesics and understand the underlying nature of general relativity.

The intended learning outcomes of this instructional framework are structured as follows:
\begin{itemize}[nosep]
    \item \textbf{LO1:} Demonstrate understanding of proper-time dynamics in the Kerr geometry by classifying trajectories using azimuthal distance.
    \item \textbf{LO2:} Apply computational skills to reproduce these numerical simulations, including extending the framework to evaluate other spacetime metrics.
    \item \textbf{LO3 (Advanced):} Analyse the conjugate points of geodesics by applying numerical Jacobi field analysis.
\end{itemize}

While the analysis in this paper has considered the Kerr metric, exactly the same mathematical framework can be applied to any metric. This broad applicability allows aspects of this framework to be used to structure assignments within a course on general relativity.

\section{Mathematical Structure}\label{sec:methods}
\subsection{Equation of Motion and Reparametrisation}\label{subsec:equationofmotion}
We use geometrised units ($G=c=1$) to focus on relative properties throughout this paper. Unless otherwise stated, the black hole mass $M$ is set to 1 to make all relevant variables (e.g. radius $r$) dimensionless ratios. Throughout this work, we adopt the metric signature $(-, +, +, +)$.

Motion along a timelike path is governed by the generalised geodesic equation. For a particle with four-velocity $u^\alpha = dx^\alpha/d\tau$ parameterised by proper time $\tau$, the equation of motion is:
\begin{equation} \label{eq:geodesic_general}
    \frac{d^2x^\alpha}{d\tau^2} + \Gamma^\alpha_{\beta\gamma}\frac{dx^\beta}{d\tau}\frac{dx^\gamma}{d\tau} = a^\alpha
\end{equation}
where $\Gamma^\alpha_{\beta\gamma}$ are Christoffel symbols and $a^\alpha=Du^\alpha/d\tau$ represents the four-acceleration.\footnote{Throughout, the plain symbol $a$ refers to the Kerr spin parameter. The four-acceleration vector and its components are represented by $a^{\alpha}$. Later, the maximal acceleration (a scalar) will be represented by $a_0$.} Note that $a^\alpha = 0$ represents a free-fall path (a geodesic). The four-velocity vector satisfies the normalisation condition:
\begin{equation} \label{eq:velocity_norm}
    g_{\alpha\beta}u^\alpha u^\beta = u^\alpha u_\alpha = -1
\end{equation}
The four-acceleration $a^\alpha$ must be orthogonal to the four-velocity $u^\alpha$. So,
\begin{equation} \label{eq:accel_ortho}
    g_{\alpha\beta}a^\alpha u^\beta = a^\alpha u_\alpha = 0
\end{equation}
In this work, we study the twin paradox in the geometry of a rotating black hole, described by the Kerr metric. In Boyer-Lindquist coordinates $(t,r,\theta,\phi)$ \cite{boyer1967maximal}, the line element is given by:
\begin{equation} 
    \begin{aligned} 
    ds^2 = & -\left(1 - \frac{2Mr}{\Sigma}\right)dt^2 - \frac{4Mar\sin^2\theta}{\Sigma}dtd\phi + \frac{\Sigma}{\Delta} dr^2 \\ 
    & + \Sigma d\theta^2 + \left(r^2 + a^2 + \frac{2Ma^2r \sin^2\theta}{\Sigma}\right)\sin^2\theta d\phi^2 
    \end{aligned} \label{eq:kerr_metric} 
\end{equation}
where $a$ is the Kerr spin parameter and the auxiliary functions $\Sigma$ and $\Delta$ are defined as:
\begin{align}
    \Sigma(r, \theta) & \equiv r^2 + a^2\cos^2\theta \label{eq:sigma_def} \\
    \Delta(r) & \equiv r^2 - 2Mr + a^2 \label{eq:delta_def}
\end{align}
Equation \ref{eq:geodesic_general} constitutes a system of ordinary differential equations in which $\tau$ is an independent variable. However, when defining a twin paradox scenario, the boundary conditions are prescribed in the Boyer-Lindquist coordinates, with their time components given in terms of coordinate time $t$, rather than proper time $\tau$. In particular, direct integration of the system with respect to $\tau$ complicates the determination of the upper integration limit, since the total proper time is not known a priori. 

To simplify the numerical procedure, we reformulate the system with coordinate time $t$ as the independent variable using the chain rule \cite{fung2016computational}. We define:
\begin{equation}
    \nu^\tau \equiv \frac{d\tau}{dt}, \qquad \nu^t \equiv \frac{dt}{dt}=1, \qquad \nu^i \equiv \frac{dx^i}{dt}
\end{equation}
where $i=r,\theta,\phi$. After substitution, the original equation of motion \ref{eq:geodesic_general} is transformed into:
\begin{align}
    \label{eq:geodesic_transformed}
    \frac{d\nu^\tau}{dt} &= \nu^\tau\Gamma^t_{\alpha\beta}\nu^\alpha\nu^\beta -a^t(\nu^\tau)^3 \\
    \frac{d\nu^i}{dt} &= \nu^i\Gamma^t_{\alpha\beta}\nu^\alpha\nu^\beta - \Gamma^i_{\alpha\beta}\nu^\alpha\nu^\beta - (a^i - a^t\nu^i)(\nu^\tau)^2
\end{align} 
Correspondingly, this reparameterisation transforms Equations \ref{eq:velocity_norm} and \ref{eq:accel_ortho} into:
\begin{align} 
    \label{eq:velocity_norm_transformed}
    g_{\alpha\beta}\nu^\alpha\nu^\beta &= -(\nu^\tau)^2\\
    \label{eq:accel_ortho_transformed}
    g_{\alpha\beta}a^\alpha\nu^\beta &= 0
\end{align}
When setting the initial state of a trajectory, the four components of the initial four-velocity or four-acceleration are not mutually independent. Owing to intrinsic constraints, only three components represent free variables. We typically take the three spatial components as independent inputs. For the four-velocity, given the three spatial components, the time component is determined via Equation \ref{eq:velocity_norm_transformed}:
\begin{equation}\label{eq:nu_tau}
     \nu^\tau =\sqrt{-g_{\alpha\beta}\nu^\alpha\nu^\beta}
\end{equation}
Similarly, for the four-acceleration, its time component is derived from Equation \ref{eq:accel_ortho_transformed}:
\begin{equation}
    \label{eq:at}
     a^t = -\frac{g_{i\alpha}a^i\nu^\alpha}{g_{t\beta}\nu^\beta}=-\frac{g_{i\alpha}a^iu^\alpha}{g_{t\beta}u^\beta}
\end{equation}


\subsection{Finding Target Trajectories}\label{subsec:opt}
\begin{figure}[htbp] 
    \centering 
    \includegraphics[width=0.7\textwidth]{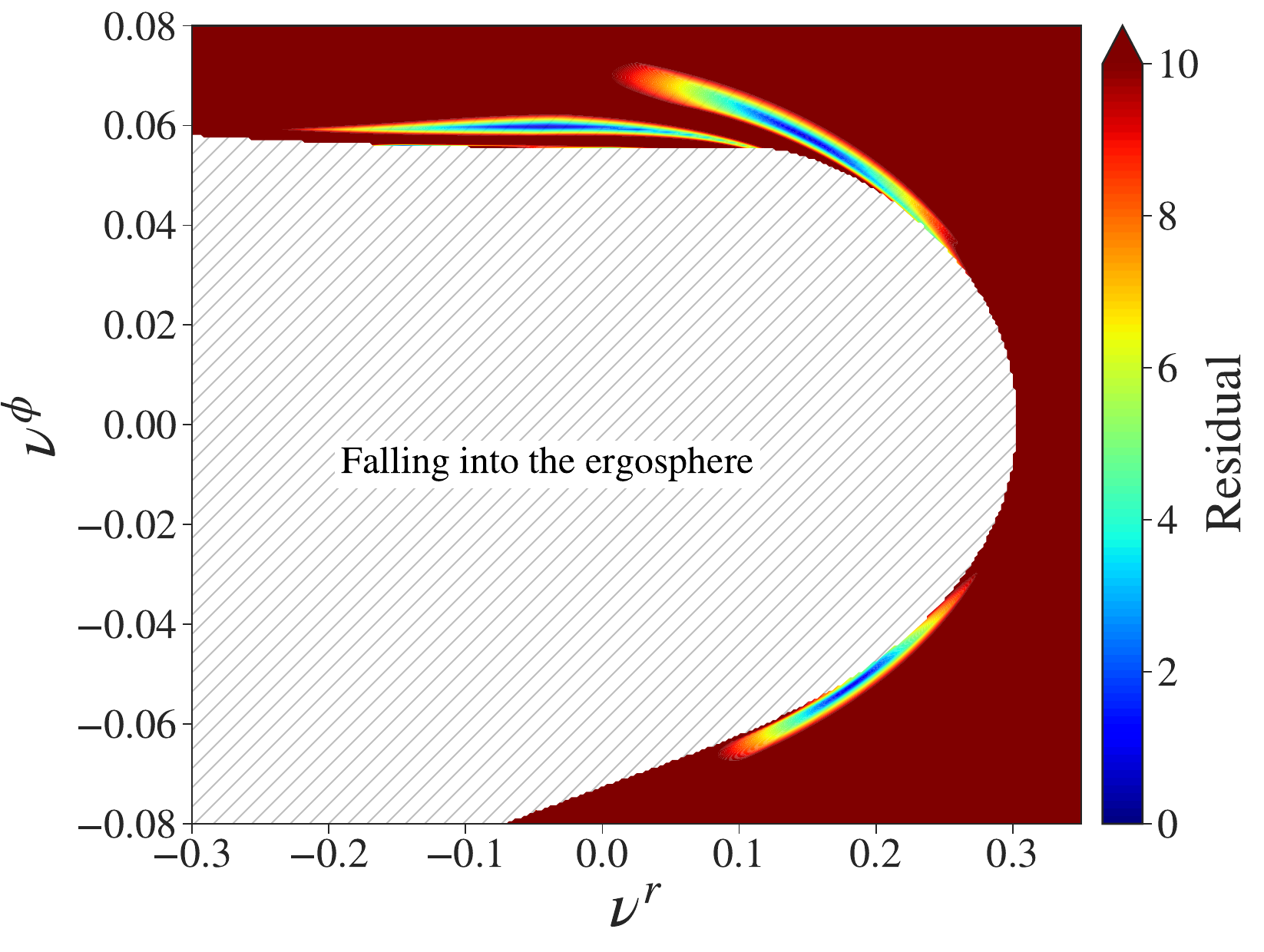} 
    \caption{Residual map for geodesic motion in the equatorial plane ($\theta = \pi/2$) with Kerr spin parameter $a=0.9$, generated over a $301 \times 301$ initial velocity grid. The boundary conditions are set to: initial event $(0, 6, \pi/2, \pi/2)$ and reunion event $(98, 6, \pi/2, 3\pi/2)$. The final coordinate time $t = 98$ corresponds to the orbital period of a prograding circular orbit at $r = 6$. The striped regions indicate initial velocities that cause trajectories to fall into the ergosphere, thus their residual values are numerically set to NaN and are therefore not rendered in the plot.}
    \label{fig:res_no_initial}
\end{figure}
To compare proper times between two prescribed events, we must first determine physically admissible trajectories that satisfy the boundary conditions. This is a shooting problem; for a given acceleration profile, determine the initial velocity such that the observer arrives at the designated spatial position at a predetermined coordinate time. We reformulate it as an optimisation problem by constructing a residual function that quantifies the deviation of a trial trajectory from the target. Specifically, this function takes a candidate initial velocity as input and returns the spatial discrepancy between the attained position and the target position at the final coordinate time. While the mathematical form of the residual function is not unique, in this work we adopt the following definition:
\begin{equation}\label{eq:residual}
     R(\nu_{\text{initial}}) =  |r_f-r_e| + r_e\cdot|\theta_f - \theta_e| + r_e\cdot\sin{\theta_e}\cdot|\phi_f - \phi_e|
\end{equation}
where $\nu_{\text{initial}}$ represents the initial velocity (with respect to coordinate time $t$) to be determined, the subscript `$f$' and `$e$' denote the final spatial position at the end of integration and the predetermined reunion position, respectively. The radial distance $r$ is some positive real number, the polar angle $\theta\in[0,\pi]$ and the azimuthal angle $\phi\in[0,2\pi)$. This residual function is constructed to be non-negative, and finding its minimum is equivalent to solving the original boundary value problem. Although this formulation applies to general 3D motion, we use the 2D equatorial case ($\nu^r - \nu^\phi$ space) in this section to illustrate the optimisation landscape, with 3D results presented in the Results section.

To reduce potential numerical instabilities, our analysis restricts trajectories to the region outside the ergosphere. If the trajectory intersects the boundary of the ergosphere, the integration is terminated. This trial is then marked as invalid by setting the corresponding residual value to NaN.\footnote{NaN stands for "Not a Number", a standard computational flag for an undefined or invalid value. Here, it acts as an infinite penalty during the optimisation process to reject invalid paths.} This is equivalent to introducing an infinitely high potential barrier, thereby preventing exploration into forbidden regions.

Figure \ref{fig:res_no_initial} is a typical residual map of equatorial motion ($\theta=\frac{\pi}{2}$ and $\nu^\theta=0$) over the initial velocity parameter space. Large striped regions correspond to initial velocities that would cause the spacecraft to fall into the ergosphere. It reveals several discrete ``valleys" of low residual values, indicating the presence of solutions in its vicinity. Pedagogically, these maps act as a visual tool, allowing students to intuitively examine the parameter space prior to precise numerical optimisation. Based on the residual map, we adopt a two-step strategy to find those solutions. The specific details have been included in Appendix \ref{secA1}.


\subsection{Numerical Jacobi Field Analysis}\label{sec:method_jacobi}
In this framework, the Jacobi equation is employed to investigate the behaviour of known geodesics (precomputed via optimisation) and is closely linked to conjugate points. The Jacobi equation is also referred to as the geodesic deviation equation in the broader literature on general relativity, and we use the former throughout this article. Consider a family of geodesics, and let $J^\mu$ be the deviation vector that connects points of equal proper time $\tau$ on infinitesimally separated geodesics within the family. The evolution of $J^\mu$ along a central geodesic $x^\alpha(\tau)$ with four-velocity $u^\alpha = dx^\alpha/d\tau$ is governed by the Jacobi Equation:
\begin{equation} 
\label{eq:jacobi}
\frac{D^2J^\mu}{d\tau^2} + R^\mu_{\ \alpha\beta\gamma}u^\alpha J^\beta u^\gamma = 0
\end{equation}
where $R^\mu_{\ \alpha\beta\gamma}$ is the Riemann curvature tensor. The vanishing of the Jacobi field along a geodesic leads to the definition of conjugate points. A conjugate point arises where a non-trivial Jacobi field vanishes at two distinct points along the timelike path (e.g. $J(0) = J(\tau_0) = 0$), marking the reconvergence of geodesics that were initially separating.

\begin{definition}[\normalfont see Section~5.5 Definition~2 in \cite{do2016differential}]\label{definition2}
Let $\gamma$ be a geodesic with $\gamma(0) = p$. We say that the point $q = \gamma(\tau_0)$ is conjugate to $p$ relative to the geodesic $\gamma$ if there exists a Jacobi field $J(\tau)$ which is not identically zero along $\gamma$ with $J(0) = J(\tau_0) = 0$.
\end{definition}
\noindent The connection between conjugate points and proper-time maximisation can be understood through an analogy with ordinary calculus. For a function $f(x)$, the condition $f'(x)=0$ identifies stationary points, which may be local maxima, local minima, or stationary points of inflection. To distinguish between them, one must examine $f''(x)$, where $f''(x)<0$ indicates a local maximum. Similarly, the geodesic condition corresponds to the vanishing of the first variation of proper time ($I'(0)=0$), which is only a necessary condition. Whether the proper time is actually maximal depends on the second variation $I''(0)$. As explained by Boyer \cite{boyer1964clock}, if a geodesic segment contains an interior conjugate point, there exists a family of variations with $I''(0)>0$, meaning the geodesic is not a local maximum. Conversely, $I''(0)$ remains negative for all sufficiently small variations in the absence of conjugate points along the segment, and the geodesic locally maximises the proper time. This is formalised in the following theorem.

\begin{figure}[htbp] 
    \centering 
    \includegraphics[width=0.7\textwidth]{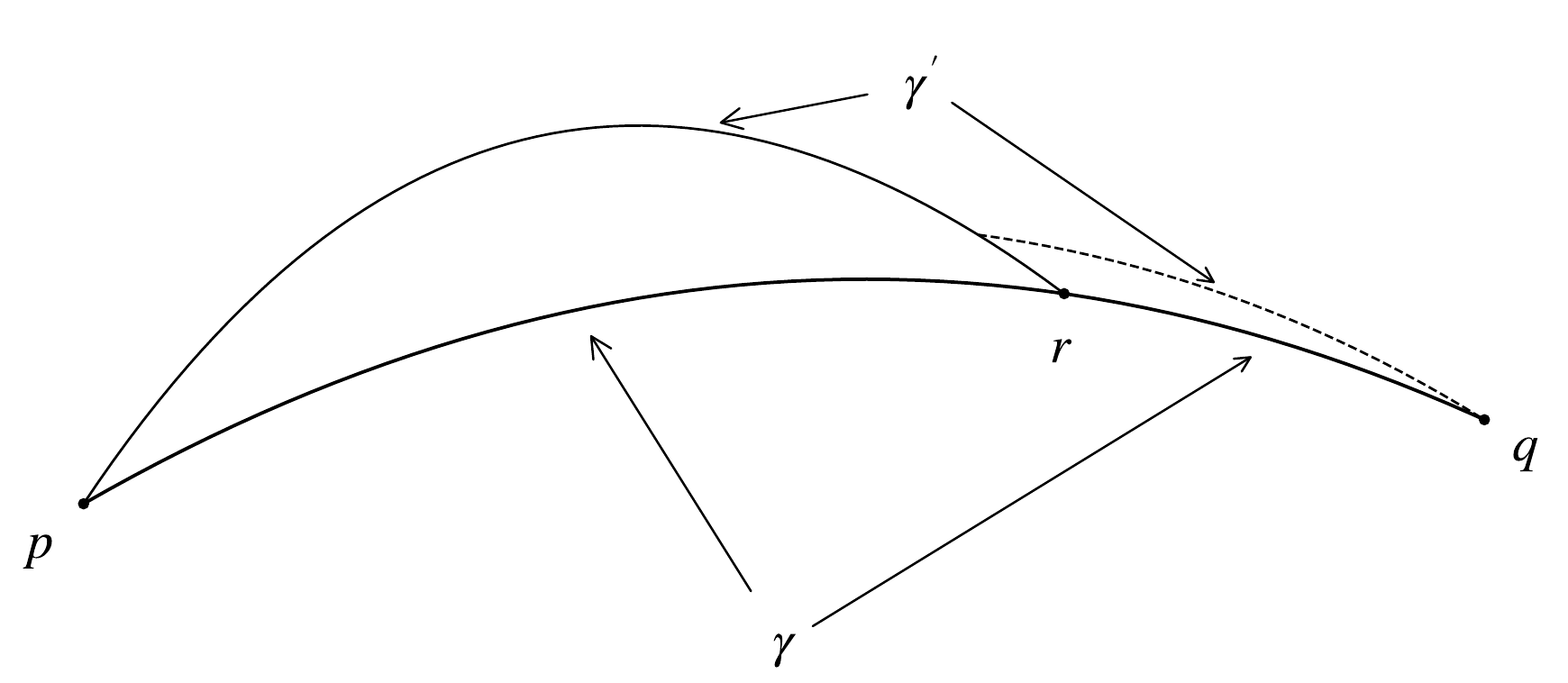} 
    \caption{Illustration of a conjugate point. The geodesic $\gamma$ passes through $p$, a conjugate point $r$, and $q$. A neighbouring timelike curve $\gamma'$ joining the same endpoints yields a greater proper time.}
    \label{fig:conjugate}
\end{figure}

\begin{theorem}[\normalfont see Theorem~9.3.3 in \cite{Wald:1984rg}]\label{theorem1}
Let $\gamma$ be a smooth timelike curve connecting two points $p$ and $q$. Then the necessary and sufficient condition that $\gamma$ locally maximise the proper time between $p$ and $q$ over smooth one parameter variations is that $\gamma$ be a geodesic with no conjugate points to p between p and q.
\end{theorem}
\noindent Theorem \ref{theorem1} identifies the absence of conjugate points as the definitive criterion for a timelike path to be a local maximum of proper time. Figure \ref{fig:conjugate} intuitively shows that if a timelike geodesic $\gamma$ connecting $p$ and $q$ contains a conjugate point $r$ between them, a nearby timelike curve $\gamma'$ connecting the same endpoints can be constructed with greater proper time. However, before passing this conjugate point $r$, the geodesic $\gamma$ is still the locally longest.

It is important to emphasise that this condition does not guarantee a global maximum. In curved spacetimes with non-trivial topology or gravitational lensing (such as near a black hole), multiple timelike paths may connect the same two events. In these cases, identifying the global maximum requires a direct comparison of the proper times ($\tau = \int \sqrt{-g_{\alpha\beta}\dot{x}^\alpha\dot{x}^\beta}d\lambda$) across all candidate timelike paths \cite{sokolowski2017geometric}. Nevertheless, the variational principle ensures that if a global maximum exists, it must be one of the geodesics.

While the analysis of conjugate points is crucial in advanced general relativity, the complexity of the Jacobi equation within the Kerr geometry makes analytical solutions intractable for coursework. Therefore, a numerical procedure is essential. To facilitate numerical integration, we rewrite the Jacobi field Equation \ref{eq:jacobi} in an equivalent form of a first-order ordinary differential equation system:
\begin{align}
    \frac{dJ^\mu}{d\tau} &= W^\mu - \Gamma^\mu_{\alpha\beta}u^\alpha J^\beta \\
    \frac{dW^\mu}{d\tau} &= -R^\mu_{\ \alpha\beta\gamma}u^\alpha J^\beta u^\gamma - \Gamma^\mu_{\alpha\beta}u^\alpha W^\beta
\end{align}
Here, $J$ is the Jacobi field and $W\equiv DJ/d\tau$ is its covariant derivative along the geodesic. The required components of the Riemann curvature tensor are provided by a precomputed numerical function. By the fundamental existence and uniqueness theorem for ordinary differential equations, a complete set of initial conditions including the initial Jacobi field $J(0)$ and its covariant derivative $W(0)$ determines a unique solution $J(\tau)$ for the system.

Direct numerical integration of a single Jacobi field is not feasible, since the definition of a conjugate point is just existential and does not prescribe how to construct its initial conditions. Solving the Jacobi equation as an initial-value problem requires specifying both $J(0)$ and $W(0)$, while one may set $J(0)=0$ according to the definition of conjugate point, the initial covariant derivative $W(0)$ remains unknown.

To overcome this problem, we numerically compute a complete basis of vectors that spans the space of all possible Jacobi fields instead of solving for a particular Jacobi field. Specifically, we construct three linearly independent Jacobi fields, $J_1(\tau)$, $J_2(\tau)$, and $J_3(\tau)$, which span the three-dimensional subspace orthogonal to the geodesic's four-velocity $u(\tau)$. The detailed construction of their initial conditions using the Gram-Schmidt process is provided in Appendix \ref{secA2}.

The search for conjugate points reduces to identifying points where these three Jacobi fields become linearly dependent. In our numerical implementation, this linear dependence is evaluated using the Gram matrix $G$ defined in Equation \ref{eq:gram}. A vanishing determinant of this matrix, $det(G)=0$, provides a necessary and sufficient condition for the linear dependence of the vector set, where $\langle ,\rangle$ denotes the inner product.
\begin{equation}\label{eq:gram}
G = 
\begin{bmatrix}
    \langle J_{1}, J_{1} \rangle & \langle J_{1}, J_{2} \rangle & \langle J_{1}, J_{3} \rangle \\
    \langle J_{2}, J_{1} \rangle & \langle J_{2}, J_{2} \rangle & \langle J_{2}, J_{3} \rangle \\
    \langle J_{3}, J_{1} \rangle & \langle J_{3}, J_{2} \rangle & \langle J_{3}, J_{3} \rangle
\end{bmatrix}
\end{equation}


\section{Results}\label{sec:results}
\subsection{Analysis of Equatorial Motion (\texorpdfstring{$\theta = \pi/2$}{θ = π/2})}
\subsubsection{Path Multiplicity and Azimuthal Distance}\label{subsec:multiplicity}
\begin{figure}[htbp] 
    \centering 
    \includegraphics[width=0.95\textwidth]{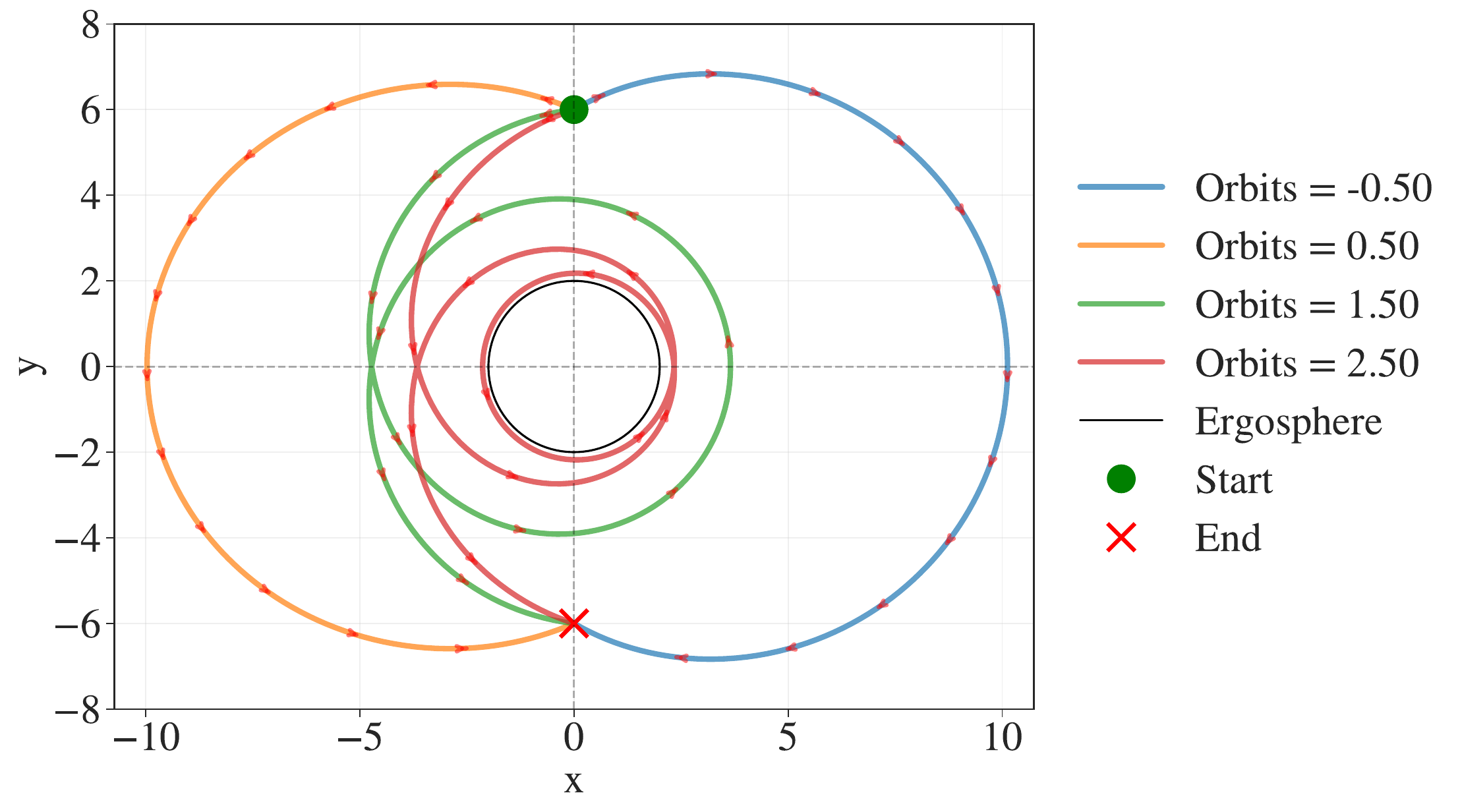} 
    \caption{Geodesics in a Kerr spacetime with Kerr spin parameter $a = 0.9$. The trajectories connect the initial event $(t, r, \theta, \phi) = (0, 6, \pi/2, \pi/2)$ (green point) and the reunion event $(98, 6, \pi/2, 3\pi/2)$ (red cross). The central black ring represents the boundary of the ergosphere. The red arrow indicates the direction of motion. The ``Orbits" label represents the total number of orbital windings, defined as the value of $\Delta\phi/2\pi$. A positive value refers to prograde motion and a negative value refers to retrograde motion. $a > 0$ defines a counterclockwise black hole rotation. Animations of these trajectories are available in the Supplementary Material.}
    \label{fig:traj_angular_geodesics}
\end{figure}

In Section \ref{subsec:opt}, we introduced a scenario of the twin paradox (see Figure \ref{fig:res_no_initial}) with a Kerr spin parameter $a=0.9$, and boundary conditions of an initial event $(t, r, \theta, \phi) = (0, 6, \pi/2, \pi/2)$ and a reunion event $(98, 6, \pi/2, 3\pi/2)$. To visually check the validity of the optimisation results, we numerically integrated the corresponding trajectories using these initial velocities. 

Figure \ref{fig:traj_angular_geodesics} illustrates the physical validity of solutions generated by the two-step algorithm. The term ``Orbits” is used as a simplified measure of the azimuthal distance of the trajectory, defined as the value of $\Delta\phi/2\pi$. A positive ``Orbits" corresponds to prograde motion (aligned with the black hole's rotation) and a negative value corresponds to retrograde motion. 

Azimuthal distance is introduced to address a methodological issue. In flat spacetime, it is straightforward to compare the proper time of the unique geodesic with that of accelerated paths, typically by plotting proper time against the magnitude of acceleration. In our case, however, all four worldlines under comparison are geodesics, each with zero acceleration.

\begin{figure}[htbp] 
    \centering 
    \includegraphics[width=0.7\textwidth]{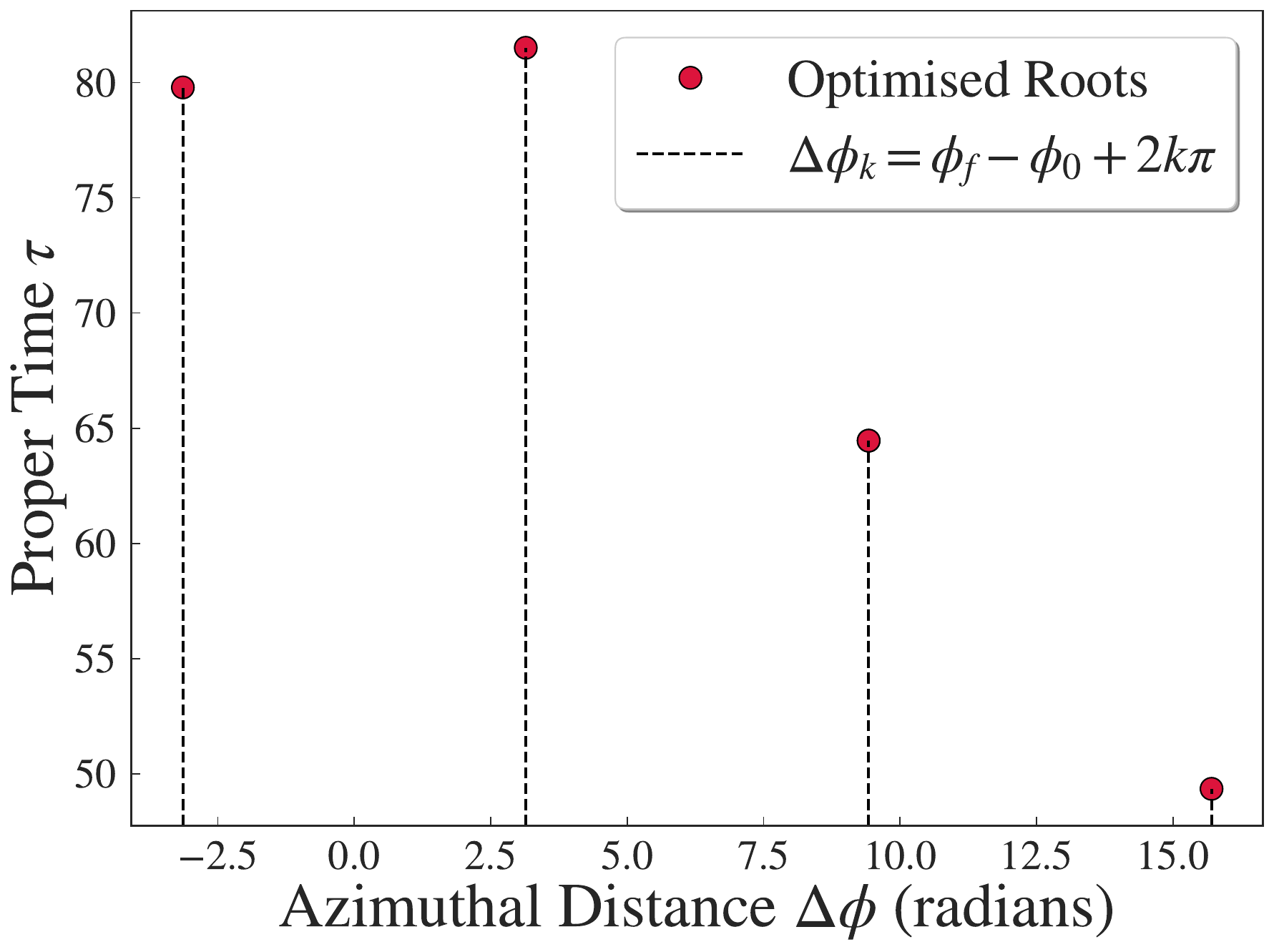} 
    \caption{Relationship between the proper time of geodesics and their azimuthal distance. The red dots represent the trajectories corresponding to the initial velocities obtained by optimisation. The proper time is negatively correlated with the magnitude of the azimuthal distance; the larger the azimuthal distance, the shorter the proper time. The azimuthal distance $\Delta\phi$ required to satisfy the boundary conditions is given by $\Delta\phi = (\phi_f - \phi_0) + 2k\pi$, where $\phi_0$ and $\phi_f$ are the azimuthal coordinates of the start and reunion events and $k$ is some integer.}
    \label{fig:traj_angular_geodesics_timevsphi}
\end{figure}

Figure \ref{fig:traj_angular_geodesics_timevsphi} demonstrates a correlation between proper time and azimuthal distance, which supports the use of the latter as a key comparative parameter in our analysis. The figure shows that, as a general trend, a greater magnitude of azimuthal distance (i.e., more orbits) corresponds to a shorter proper time (with a minor deviation at $\Delta\phi = \pm\pi$ to be discussed below). 

The choice of azimuthal distance arises from its parametric simplicity and physical interpretability. Since the black hole mass $M$ and Kerr spin parameter $a$ remain fixed, and our objective is to identify an alternative indicator to acceleration, the most natural candidates are therefore the spacetime coordinates. With the proper time already designated as the dependent variable, the azimuthal angle $\phi$ offers a distinct advantage among the three spatial coordinates ($r$, $\theta$, $\phi$) because its cumulative change is generally monotonic due to the axial symmetry of the Kerr metric. The net change $\Delta\phi$ must satisfy the relation $\Delta\phi = (\phi_f - \phi_0) + 2k\pi$, where $\phi_0$ and $\phi_f$ refer to the azimuthal coordinates of the start and reunion events and $k$ is some integer. In contrast, motion along the $r$ and $\theta$ directions typically oscillates within specific intervals. While other parameters, such as some indicator related to the Carter constant $Q$ \cite{carter1968global}, could also serve this purpose, we use $\Delta\phi$ for its intuitive geometric meaning.

Figure \ref{fig:traj_angular_geodesics_timevsphi} also reveals another phenomenon in the comparison of two trajectories with identical absolute azimuthal distance ($|\Delta\phi| = \pi$). The two data points on the left correspond to $\Delta\phi = -\pi$ (retrograde) and $\Delta\phi = +\pi$ (prograde), respectively. Although the magnitudes of their azimuthal distances are equal, the data indicate a slight difference in proper time. The prograde trajectory exhibits a marginally longer proper time than the retrograde trajectory. This asymmetry can be attributed to the frame-dragging effect induced by the black hole's rotation. We further investigate whether these observations also apply to accelerated trajectories in the following section.

\subsubsection{Azimuthally Accelerated Trajectories}\label{subsec:azimuthal}
The optimisation algorithm established in Section \ref{subsec:opt} can also be applied to determine accelerated trajectories. However, unlike geodesics, the solution of an accelerated trajectory requires both direction and the evolution of the magnitude of the four-acceleration $a^\mu$. In this section, to simplify the model and facilitate comparison with the geodesic results obtained in Section \ref{subsec:multiplicity}, we restrict the motion to the equatorial plane ($\theta = \pi/2$) with purely azimuthal acceleration ($a^r = a^\theta = 0$, $a^\phi \neq 0$). The time component $a^t$ of the four-acceleration is derived from the four-velocity normalisation condition (see Equation \ref{eq:at}). We adopt a piecewise-linear acceleration profile following Fung et al. \cite{fung2016computational}, where $a^\phi = 4a_0 /T\cdot|t-T/2|-a_0$. Here, $a_0$ is the maximum acceleration and $T$ is the reunion time.

\begin{figure}[htbp]
    \centering
    \includegraphics[width=0.95\textwidth]{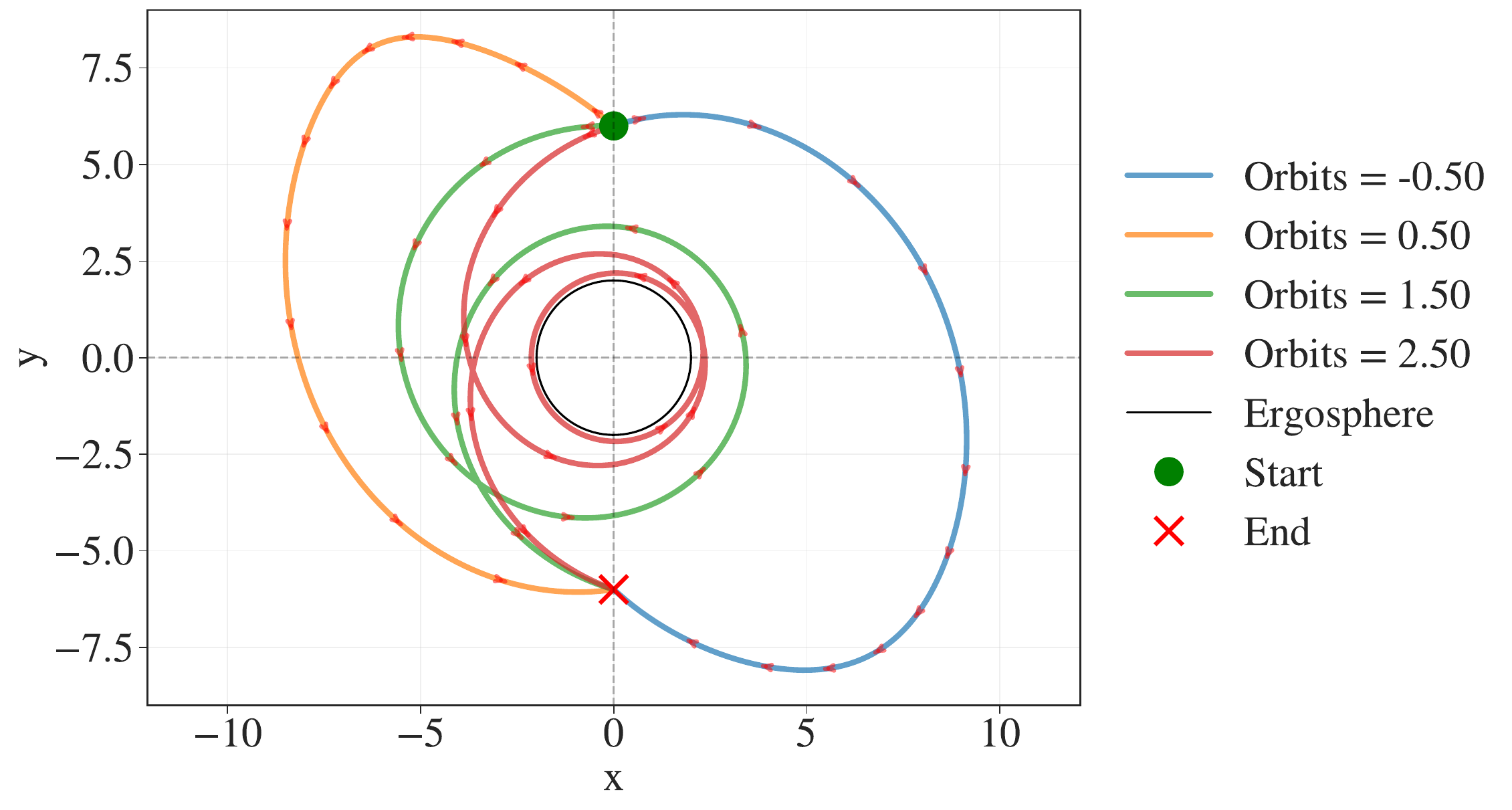}
    \caption{The accelerated trajectories in the equatorial plane with a maximal acceleration $a_0=-0.001$. The azimuthal acceleration follows a piecewise-linear function of coordinate time $t$, expressed as $4a_0 /T\cdot|t-T/2|-a_0$, where $T$ is the total coordinate time determined by the reunion event.}
    \label{fig:linearacc_res-0.001}
\end{figure}

We use the case with the maximal acceleration $a_0 = -0.001$ as an example. All other parameter settings are kept identical to the geodesic case studied in Section \ref{subsec:multiplicity}. The resulting trajectories are shown in Figure \ref{fig:linearacc_res-0.001}. By repeating the same analysis, we generated multiple sets of trajectories with purely azimuthal acceleration and stored the resulting data for subsequent analysis. All trajectories share the piecewise-linear acceleration profile but vary in their maximum acceleration $a_0$.

\begin{figure}[htbp]
    \centering
    \begin{minipage}[b]{0.48\textwidth}
        \centering
        \includegraphics[width=\textwidth]{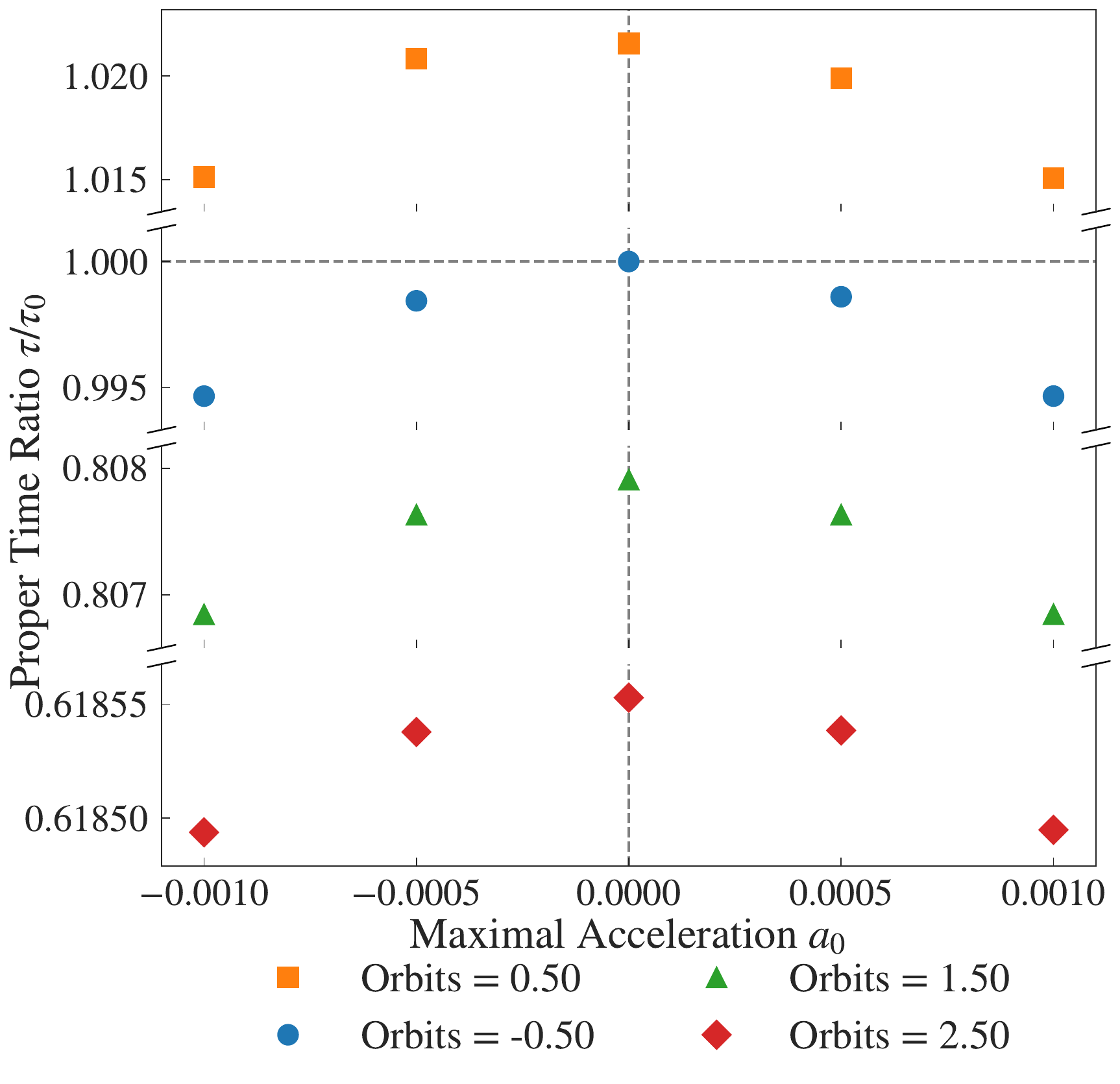}
    \end{minipage}
    \hspace{0.01\textwidth}
    \begin{minipage}[b]{0.48\textwidth}
        \centering
        \includegraphics[width=\textwidth]{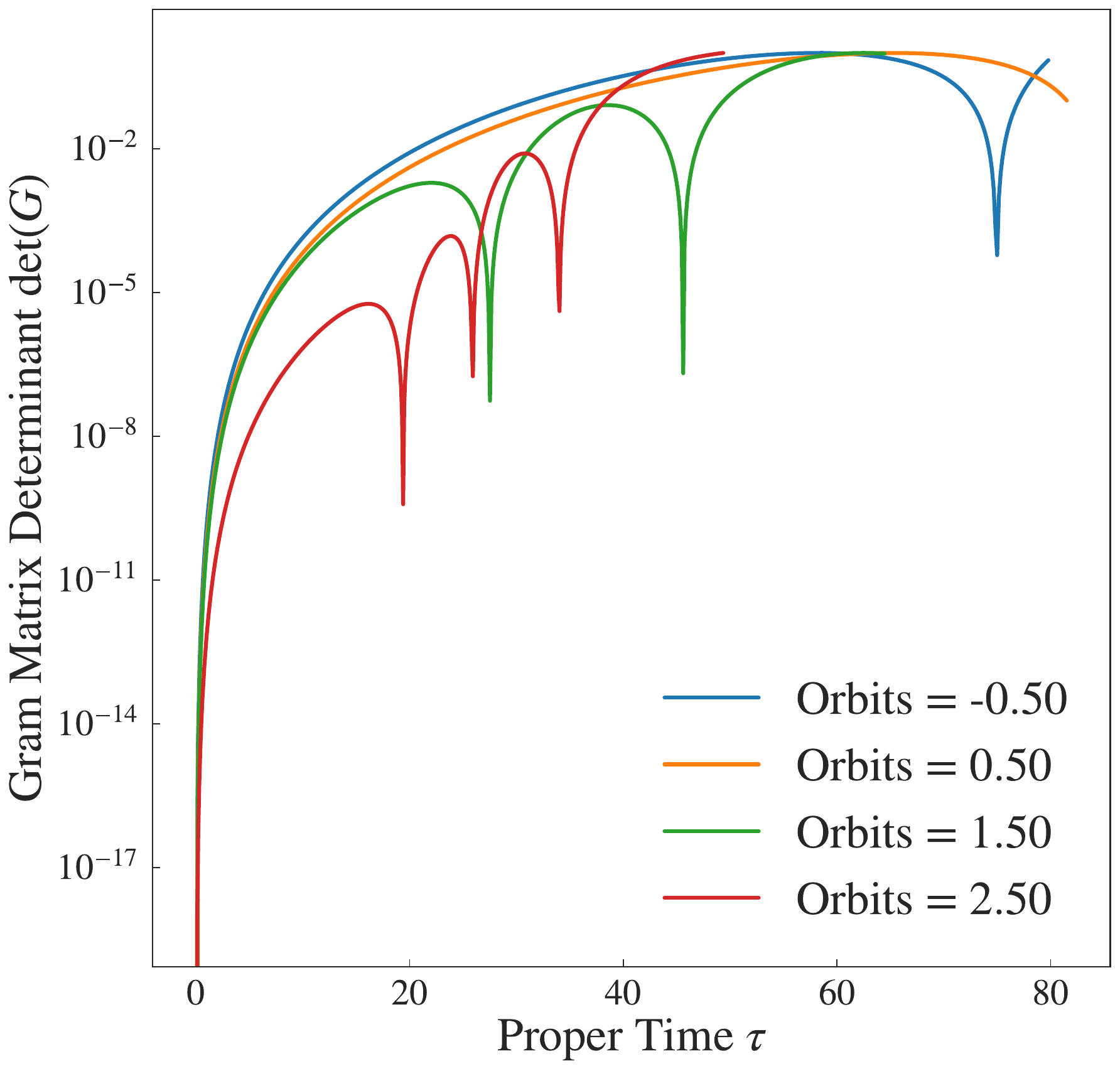}
    \end{minipage}
    \caption{Left: Normalised proper time versus maximum acceleration $a_0$ across different azimuthal distances $\Delta\phi$. Right: Evolution of the Gram matrix determinant with proper time. The determinant is normalised to its maximum value. Local minima (cusps) in the curve indicate conjugate points. The results show that only the trajectory with Orbits $= 0.5$, the prograde geodesic with the smallest azimuthal distance, contains no conjugate points.}
    \label{fig:angular_group}
\end{figure}

To isolate the effect of acceleration, we compared trajectories with the same azimuthal distance but different maximal accelerations. Representative trajectories were selected from the stored dataset, and the relationship between their proper time and maximal acceleration ($a_0$) is plotted in Figure~\ref{fig:angular_group} (left), which brings together the data from all azimuthal distance groups into a single plot. In this plot, the proper times of all trajectories are normalised against the proper time of the geodesic from the Orbits $= -0.5$ group. Thus, the vertical axis represents the ratio of the proper time of each trajectory to that of this specific geodesic. 

The plot can be interpreted in two distinct ways: vertically (fixed $a_0$), for trajectories with identical acceleration, higher Orbits exhibit shorter proper time, which also shows the proper time asymmetry between prograde and retrograde motions. Based on this observation, we formulate the first two criteria:

\begin{criteria}\label{criteria1}
    For trajectories connecting the same boundary events with identical acceleration, a greater magnitude of azimuthal distance $\Delta\phi$ corresponds to a shorter proper time $\tau$.
\end{criteria}

\begin{criteria}\label{criteria2}
    For trajectories connecting the same boundary events with identical acceleration and magnitude of azimuthal distance $|\Delta\phi|$, the proper time $\tau$ of a prograde trajectory exceeds that of a retrograde trajectory.
\end{criteria}

\noindent Horizontally (fixed $\Delta\phi$), for trajectories with identical azimuthal distance, a greater acceleration magnitude yields a shorter proper time. In analogy with the simple conclusion in flat spacetime, we similarly propose for the Kerr geometry:

\begin{criteria}\label{criteria3}
    For trajectories connecting the same boundary events with identical azimuthal distance $\Delta\phi$ and acceleration shape, a greater magnitude of acceleration corresponds to a shorter proper time $\tau$.
\end{criteria}

\noindent Criterion \ref{criteria3}, however, is not a general criterion and applies only under specific conditions. Proper time is affected by the entire history of the four-acceleration, including both its direction and how its magnitude changes over time. This means no simple, universal relationship exists that connects proper time only to the acceleration's magnitude. The ``acceleration shape" condition in Criterion \ref{criteria3} is key to this constraint, as the criterion is valid only for a family of trajectories that share the same acceleration shape (i.e., direction and temporal evolution) but differ in their overall magnitude. However, since geodesics always have zero acceleration, we can still draw one key corollary:

\begin{criteria}\label{criteria4}
    For an accelerated trajectory and a geodesic connecting the same boundary events with the same azimuthal distance, the geodesic has the longer proper time.
\end{criteria}

\noindent We then perform a Jacobi field analysis on the four geodesics to locate their conjugate points. As shown in Figure \ref{fig:angular_group} (left), for trajectories with the same azimuthal distance, the proper time of an accelerated trajectory is less than that of the corresponding geodesic. It therefore suffices to analyse only the geodesics in order to determine the overall ordering of proper times among different azimuthal-distance groups.

Following the discussion in Section \ref{sec:method_jacobi}, we use proper time $\tau$ rather than coordinate time $t$ as the variable. In our approach, we first integrate the timelike paths to determine its total proper time. Using this value as the upper integration limit, we then numerically solve both the geodesic and Jacobi equations simultaneously, which avoids reparameterisation of the original Jacobi equation. The vertical axis of Figure~\ref{fig:angular_group} (right) shows the determinant of the Gram matrix (see Equation \ref{eq:gram}), which measures the linear independence of the three Jacobi fields $J_1$, $J_2$, $J_3$. The determinant is normalised to its maximum value, so it ranges between 0 and 1. Due to numerical limitations, conjugate points appear as sharp local minima rather than strict zeros. These minima correspond to points where neighbouring timelike paths begin to converge after diverging.

Figure \ref{fig:angular_group} (right) shows that only the trajectory with Orbits $= 0.5$, the prograde geodesic with the smallest azimuthal distance has no conjugate points, while all other geodesics analysed contain at least one. This finding aligns with the four previously established criteria and the proper time ordering shown in Figure \ref{fig:angular_group} (left). It also highlights a subtle aspect of the notion of ``locality”. We must distinguish between the “local maximal proper time” guaranteed by Theorem \ref{theorem1} and the “locality” implied by grouping trajectories according to azimuthal distance. If fixing $\Delta\phi$ created a ``local neighbourhood” as defined in Theorem \ref{theorem1}, then the geodesic in that group would have to be locally maximal and should therefore have no conjugate points. Yet this is not the case.

To ensure the validity of the numerical integration, we tested both the normalisation of the geodesic four-velocity and orthogonality between the Jacobi fields and the four-velocity. Both conditions are satisfied throughout the integration, which confirms the reliability of our results. Although the numerical precision prevents these quantities from being exactly constant, the deviations were found to be small ($\lesssim 10^{-7}$). 

We also note that the present Jacobi field analysis serves primarily as a qualitative tool, determining the presence or absence of conjugate points, rather than supporting quantitative conclusions. The specific number of conjugate points does not appear to carry additional physical meaning in the context of the twin paradox problem studied here. Nevertheless, we observe that although Theorem \ref{theorem1} only ensures local optimality, the geodesic with the globally longest proper time among all computed trajectories is indeed the only one free of conjugate points.

\subsection{Extension to 3D Motion: Polar Acceleration}\label{subsec:polar}
\begin{figure}[htbp]
    \centering
    \includegraphics[width=0.7\textwidth]{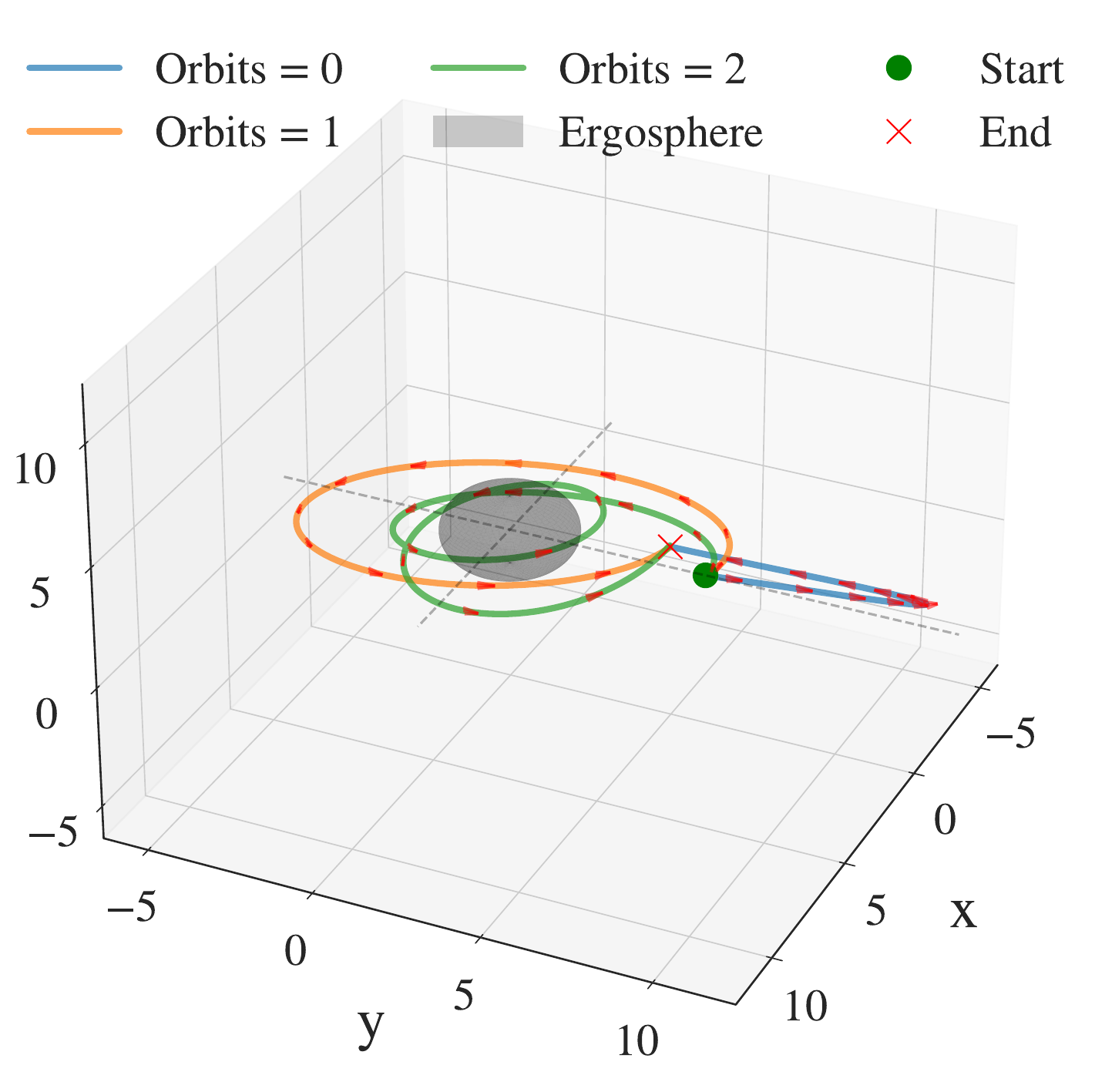}
    \caption{Geodesics including polar motion in Kerr spacetime with $a = 0.9$, sharing the same initial event $(0, 6, \pi/2, \pi/2)$ (green point) and reunion event $(98, 5, 4\pi/9, \pi/2)$ (red cross). The polar acceleration follows a sinusoidal acceleration $a_0 \sin(4\pi t/T)$, where $T$ is the total coordinate time determined by the reunion event. Animations of these trajectories are available in the Supplementary Material.}
    \label{fig:polar_separate}
\end{figure}

\begin{figure}[htbp]
    \centering
    \begin{minipage}[b]{0.49\textwidth}
        \centering
        \includegraphics[width=\textwidth]{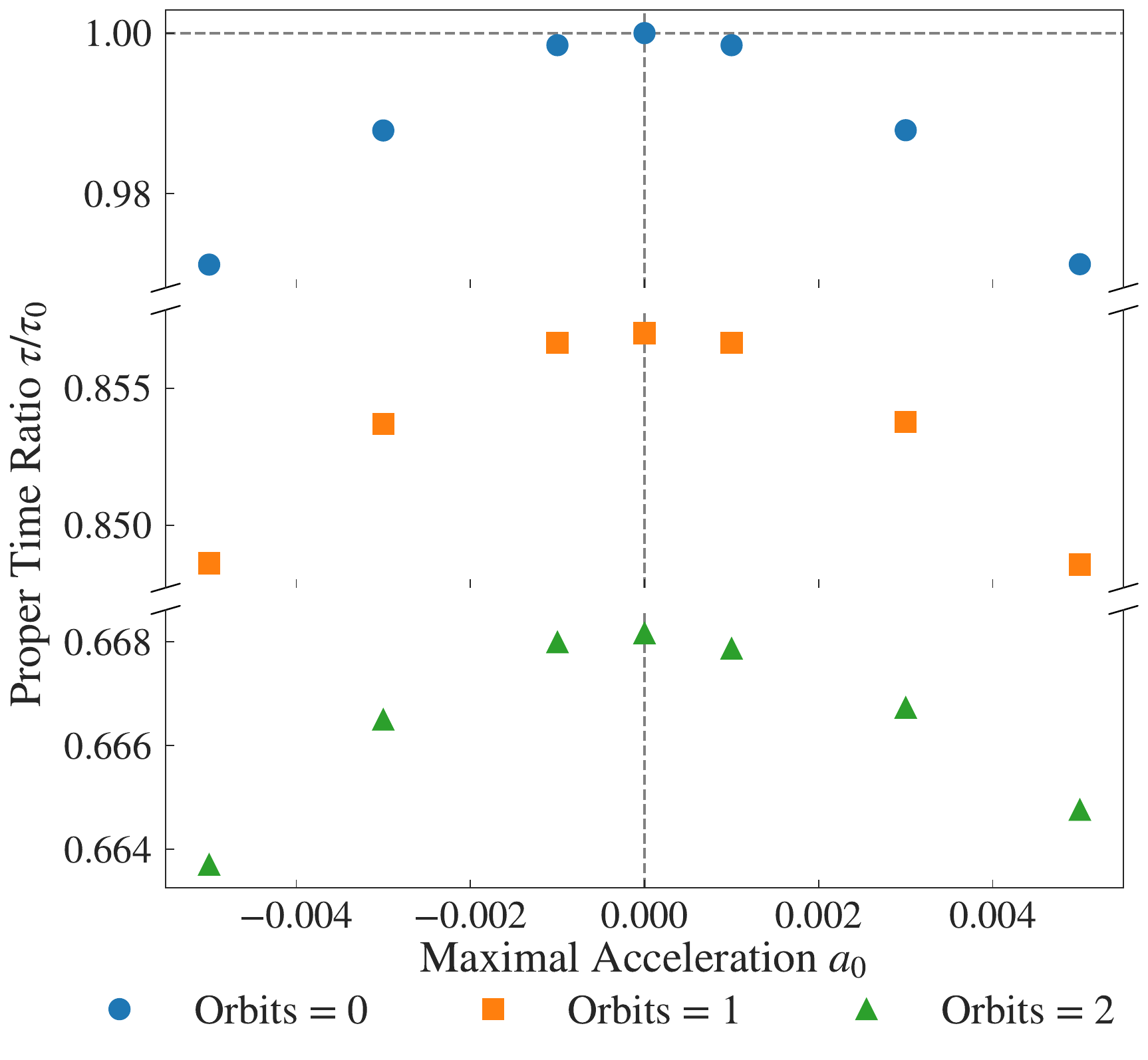} 
    \end{minipage}
    \hspace{0.0\textwidth}
    \begin{minipage}[b]{0.49\textwidth}
        \centering
        \includegraphics[width=\textwidth]{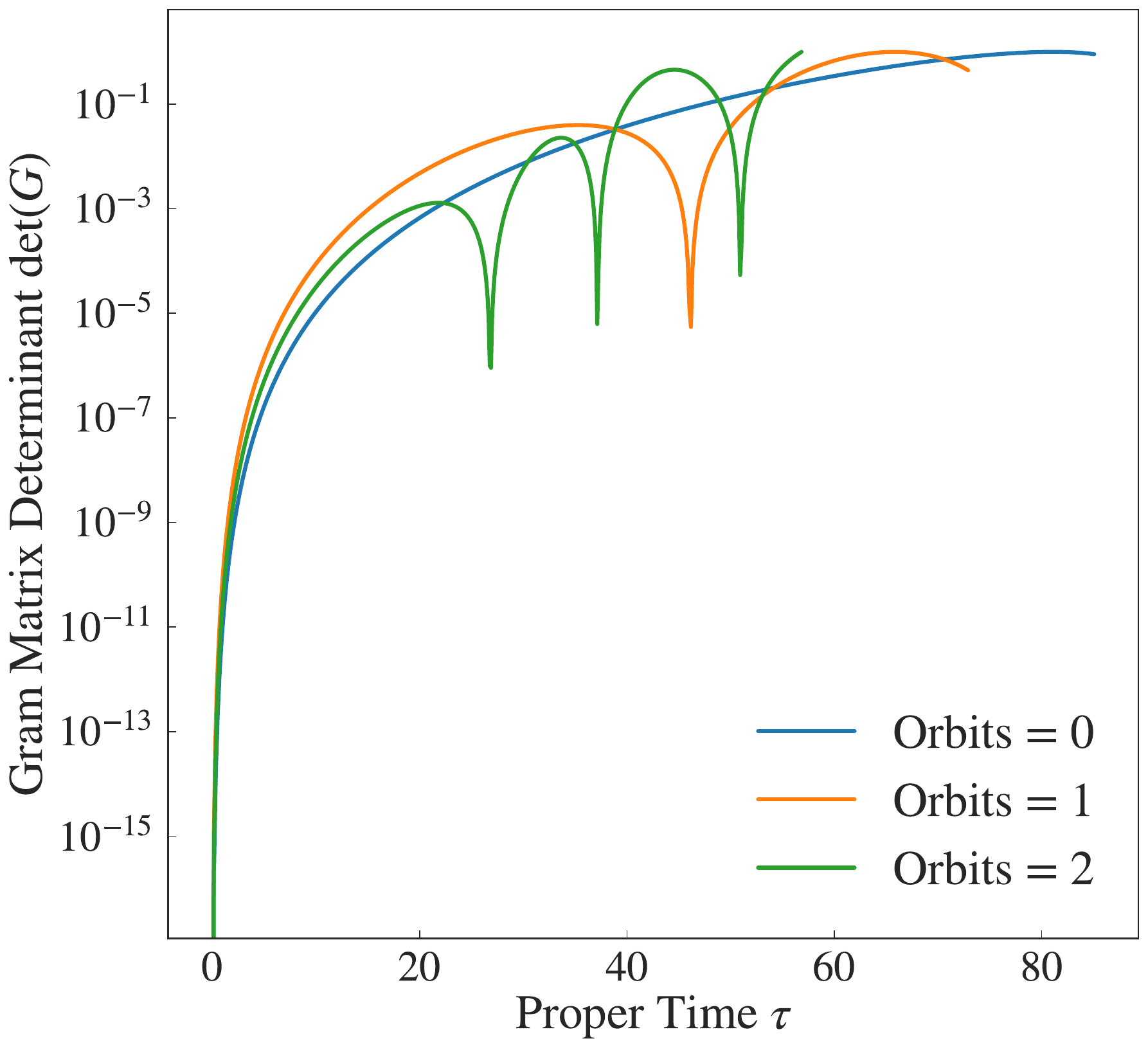}
    \end{minipage}
    \caption{Left: Normalised proper time versus maximum acceleration $a_0$ (polar acceleration). The vertical axis gives the ratio of the proper time of each trajectory to that of the geodesic ($a_0 = 0$) in the Orbits $= 0$ group. The overall trend is consistent with the concluded criteria. Right: Evolution of the Gram matrix determinant for geodesics with polar motion. The results show that the trajectory with Orbits $= 0$, the geodesic with the smallest azimuthal distance, contains no conjugate points, which is consistent with the pattern observed in Section \ref{subsec:azimuthal}.}
    \label{fig:polar_group+jacobi}
\end{figure}

As a further extension, we now apply the same methodology to the scenarios involving polar motion. In this section, the Kerr spin parameter remains $a = 0.9$ and the initial event is again $(0, 6, \pi/2, \pi/2)$. To introduce polar motion, the reunion event is chosen off the equatorial plane as $(98, 5, 4\pi/9, \pi/2)$. The acceleration is constrained to the polar direction only (i.e., $a^r = a^\phi = 0$, $a^\theta \neq 0$). Unlike the piecewise-linear profile used earlier, we adopt a sinusoidal acceleration form here, defined as $a^\theta = a_0 \sin(4\pi t/T)$. Figure \ref{fig:polar_separate} displays the corresponding geodesics obtained from the solution. We note that, due to the altered reunion position, only three distinct solution families appear in this case. Compared with the four found previously, the Orbits$= -1$ family is absent. Figure \ref{fig:polar_group+jacobi} (left) summarises the relationship between proper time, azimuthal distance and maximum acceleration in a combined plot. The results clearly show that even with the inclusion of polar motion, the behaviour remains consistent with the conclusions drawn from the equatorial case. Specifically, the proper time still decreases with both increasing maximal acceleration $a_0$ and increasing azimuthal distance.

We also performed Jacobi field analysis on these three geodesics under the polar motion scenario. Figure \ref{fig:polar_group+jacobi} (right) shows that only the Orbits $= 0$ geodesic, corresponding to the shortest azimuthal distance, exhibits no conjugate point. This result is consistent with the behaviour observed in the equatorial motion in Section \ref{subsec:azimuthal}.

\section{Pedagogical Implications}\label{sec:implications}
The computational framework and visual results presented in this paper provide a concrete summary of the twin paradox and direct visualisation of timelike path behaviour in the vicinity of rotating black holes. When relevant topics are covered in general relativity courses, these results can be directly used to support different stages of the curriculum.

At the introductory level, the standard twin paradox in flat spacetime can first be used as a hook. Then, a video demonstrating that multiple geodesics connecting the same two events can exist in the Kerr spacetime (e.g., using the animations provided in the Supplementary Material) can be shown to students. This helps students visually recognise that acceleration alone is insufficient to determine age differences. This early introduction to curved spacetime complexities breaks flat-spacetime intuitions and motivates the subsequent study of general relativity.

As the course progresses to black hole metrics, students can first be prompted to identify alternative parameters that might differentiate the proper times of multiple trajectories, with a hint to consider conserved quantities. The proper time distributions from Section \ref{sec:results} can then be introduced for a data-driven inquiry. Guiding students to extract the four criteria allows them to observe the impact of azimuthal distance and its connection to angular momentum conservation. Additionally, Criterion \ref{criteria2} provides a direct visualisation of the asymmetry induced by the frame-dragging effect.

At this stage, beyond the direct observation of results, students are encouraged to independently explore the numerical optimisation and data analysis across various spacetime metrics through computational assignments based on the Python package provided in the Supplementary Material. However, as the numerical framework employs parallel processing, minor structural adjustments may be necessary depending on the local execution environment, particularly when running interactive notebooks natively on Windows. It should be noted that the 3D scenario with polar motion is not suitable for computational assignments. The high-precision optimisation required for the 3D case often demands hours of runtime depending on local hardware, while the 2D equatorial plane case reduces this time to a matter of minutes.

For advanced or graduate coursework on Jacobi fields and conjugate points, the topic can begin with an exercise, requiring students to analytically compute Jacobi fields and locate conjugate points for a simple scenario like a circular orbit within the Schwarzschild geometry. The equations for general geodesics can then be presented to reveal the extreme difficulty of finding analytical solutions in curved spacetimes. This introduces the importance of numerical methods in modern physics. To address this, the numerical evolution of the Gram matrix determinant from Section \ref{sec:results} can be provided as a case study for rotating black holes. By tracking the roots of the determinant to count the number of conjugate points along different trajectories, students can verify Theorem \ref{theorem1}, observing the relationship between the absence of conjugate points and the maximisation of proper time.


\section{Conclusion}\label{sec:conclusion}
The twin paradox has long served as an educational tool for introducing and exploring relativistic concepts. Although it has been widely investigated in flat spacetime, its application in the vicinity of rotating black holes has remained underexplored. This study addresses this gap by extending the twin paradox analysis to the spacetime of rotating black holes.

Building upon the work of Fung et al. \cite{fung2016computational}, we refined the algorithm to handle trajectory computation in Kerr spacetime and developed visual tools suitable for educational use. We established four specific criteria showing that proper time is negatively correlated with both acceleration and azimuthal distance. To substantiate this geometric intuition, we incorporated numerical Jacobi field analysis into this context, which demonstrates that only the geodesic with the minimal azimuthal distance remains free of conjugate points in Kerr spacetime.

Our numerical and visual contributions provide an intuitive resource for general relativity courses and a significant methodological foundation for analysing the twin paradox in curved spacetime. By explicitly addressing the twin paradox within the Kerr geometry, this work helps students overcome flat-spacetime intuitions and connect abstract differential geometry theorems with numerical computation.


\ack{We thank Mali Land-Strykowski and Oliver Oayda for their helpful comments and suggestions on the manuscript.}

\roles{All authors contributed to the study conception and design. Geraint F. Lewis proposed the original idea and supervised the project. Shuiquan Bai performed the simulations and data analysis. The first draft of the manuscript was written by Shuiquan Bai, and all authors commented on previous versions of the manuscript. All authors read and approved the final manuscript.}

\data{The data points supporting the findings of this study are presented in the plots within the article. The raw data is available from the corresponding author on reasonable request.}

\begin{appendices}

\section{Optimisation Strategy}\label{secA1}
We implement our numerical framework in Python and employ Scipy \verb|minimize| \cite{2020SciPy-NMeth}, adopting a two-step strategy:
\begin{enumerate}[nosep]
  \item \textbf{Global Exploration}: A grid search based on filtered residual map is performed over the parameter space to identify multiple candidates that may contain potential optimal solutions.
  \item \textbf{Local Optimisation}: Each candidate point identified in the first stage is used as an initial guess for the \verb|minimize| function, enabling independent local optimisation.
\end{enumerate}

In the global exploration stage, we first generate a raw residual map (e.g. Figure \ref{fig:res_no_initial}) and apply a threshold (here set to 10) to isolate low-residual regions. These regions are then smoothed using a Gaussian filter. Subsequently, we apply a minimum filter to the smoothed residual map, which replaces each pixel value with the minimum value within its neighbourhood. 

By comparing the maps before and after filtering, we can then identify some candidate local minima. The logic behind this is that if a point remains unchanged after minimum filtering, it must be the minimum within its neighbourhood. We then conduct a final screening with a stricter threshold (here set to 3) to keep only sufficiently deep minima. 

The effectiveness of the global exploration is dependent on the sampling resolution of the initial residual map. If the grid is too sparse, some narrow ``solution valleys" may lie between grid points and thus escape detection. For the example of two-dimensional equatorial motion presented, we employed a $301 \times 301$ grid spanning the search window $\nu^r \in [-0.30, 0.35]$ and $\nu^\phi \in [-0.08, 0.08]$. This yields a step size of approximately $\sim 10^{-3}$ in the radial direction and $\sim 10^{-4}$ in the azimuthal direction, a resolution we found to be sufficient for capturing relevant solution regions. This proposed algorithm can be straightforwardly extended to the 3D cases including polar motion.

In the local optimisation stage, each candidate point is used to initialise the local optimiser for precise refinement. Since not all points necessarily converge to a global minimum, we discard solutions exceeding a strict residual threshold (here set to 0.01).

\section{Initialisation of Jacobi Fields}\label{secA2}
At $\tau=0$, we construct a local orthonormal frame $\{e_{\alpha}\}$. $e_{0}$ is chosen to align with the initial four-velocity. The remaining three vectors $e_{i}$ $(i=1,2,3)$ are constructed by applying the Gram–Schmidt process \ref{eq:gram_schmidt} \cite{golub2013matrix}. The set of standard coordinate basis vectors is chosen, i.e., $v_{1}=(0,1,0,0)$, $v_{2}=(0,0,1,0)$, and $v_{3}=(0,0,0,1)$.

\begin{equation}
\label{eq:gram_schmidt}
\left\{
\begin{aligned}
    e_{0} &= \frac{u}{\sqrt{|\langle u, u \rangle|}} \\
    v'_{i} &= v_{i} - \sum_{j=0}^{i-1} \langle v_{i}, e_{j} \rangle e_{j} \rlap{\quad (\text{for } i=1, 2, 3)} \\
    e_{i} &= \frac{v'_{i}}{\sqrt{|\langle v'_{i}, v'_{i} \rangle|}} \rlap{\quad (\text{for } i=1, 2, 3)}
\end{aligned}
\right.
\end{equation}
We are only concerned with the basis ${e_{1}, e_{2}, e_{3}}$. Following the Gram–Schmidt process, we then assign initial conditions to the basis vectors ${e_{1}, e_{2}, e_{3}}$ and evolve them along the geodesic to get three Jacobi fields $J_{1}$, $J_{2}$ and $J_{3}$. The initial conditions are then as follows:
\begin{equation}
    \label{eq:jacobi_initial}
    J_{i}(0) = 0 \quad \text{and} \quad W_{i}(0) \equiv \frac{DJ_i}{d\tau}(0) = e_{i} \quad \text{for } i=1,2,3
\end{equation}
Since the Jacobi equation is a linear, second order ordinary differential equation, this guarantees these three Jacobi fields are also linearly independent in a neighbourhood of the start point. The search for conjugate points thus becomes identifying where at least two of $J_{1}, J_{2}, J_{3}$ become linearly dependent. If there exists a conjugate point at some $\tau_c > 0$, then we can find some set of constants $(c_1, c_2, c_3)$ such that the linear combination $\sum_{i=1}^3 c_i J_{i}(\tau_c) = 0$. By linearity, the vector field $J(\tau) \equiv \sum c_i J_{i}(\tau)$ is also a Jacobi field. Since $J(0) = \sum c_i J_{i}(0) = 0$, we have constructed a non-trivial Jacobi field vanishing at both $\tau = 0$ and $\tau = \tau_c$, which means these points are conjugate points.

\section{Computational Implementation}\label{secA3}
The core code used in this paper was packaged into \texttt{kerrtwin}, a Python package written for the twin paradox analysis in Kerr spacetime. The code and a walkthrough Jupyter notebook (\texttt{usage\_examples.ipynb}) are provided as supplementary material for students to independently perform numerical calculation and visualisation. We also provide the original notebooks in plain script form for readers who prefer to work with the code directly.

The package is built around five components:

\begin{itemize}
  \item \texttt{KerrSpacetime}: stores the black hole parameters ($M$, $a$) and provides the metric tensor, horizon and ergosphere radii, and the timelike normalisation constraint.
  
  \item \texttt{Worldline}: integrates the equations of motion with \texttt{scipy.integrate.solve\_ivp}. A constant proper acceleration $a_0$ in a fixed spatial direction can be switched on.
  
  \item \texttt{ResidualMap}: evaluates the residual Equation \ref{eq:residual} over a grid of initial velocities $(u^r, u^\phi)$, producing a residual map as shown in this paper.
  
  \item \texttt{Optimizer}: optimise the candidates from the residual maps, subject to the timelike constraint $g_{\mu\nu}u^\mu u^\nu = -1$.
  
  \item \texttt{JacobiAnalyzer}: integrates the Jacobi equation along given geodesics. The Christoffel symbols and Riemann tensor are computed symbolically with \texttt{SymPy} and evaluated numerically via \texttt{lambdify}. The Gram determinant $\det(G)$ of three Jacobi fields is tracked to locate conjugate points.
\end{itemize}

The current implementation is restricted to the equatorial case ($\theta = \pi/2$), where $u^\theta = 0$ and the velocity parameter space is two-dimensional. Extending to off-equatorial orbits requires scanning the additional $u^\theta$ direction in the residual map. All other required packages have been listed in the file \texttt{requirements.txt}.

\end{appendices}

\setlength{\bibsep}{0pt}
\small
\bibliography{bibliography}

\end{document}